# Visualizing Uniaxial-strain Manipulation of Antiferromagnetic Domains in $Fe_{1+y}Te$ Using a Spin-polarized Scanning Tunneling Microscope

Mariam Kavai[1], Ioannis Giannakis[1], Justin Leshen[1], Joel Friedman[1], Paweł Zajdel[2] and Pegor Aynajian[1]

[1]Department of Physics, Applied Physics and Astronomy, Binghamton University, Binghamton, NY, USA
[2]Institute of Physics, University of Silesia, Chorzów, Poland

Correspondence should be addressed to: paynajia@binghamton.edu

**The quest to understand correlated electronic systems has pushed the frontiers of experimental measurements toward the development of new experimental techniques and methodologies. Here we use a novel home-built uniaxial-strain device integrated into our variable temperature scanning tunneling microscope that enables us to controllably manipulate in-plane uniaxial strain in samples and probe their electronic response at the atomic scale. Using scanning tunneling microscopy (STM) with spin-polarization techniques, we visualize antiferromagnetic (AFM) domains and their atomic structure in $Fe_{1+y}Te$ samples, the parent compound of iron-based superconductors, and demonstrate how these domains respond to applied uniaxial strain. We observe the bidirectional AFM domains in the unstrained sample, with an average domain size of ~50–150 nm, to transition into a single unidirectional domain under applied uniaxial strain. The findings presented here open a new direction to utilize a valuable tuning parameter in STM, as well as other spectroscopic techniques, both for tuning the electronic properties as for inducing symmetry breaking in quantum material systems.**

High-temperature superconductivity in cuprates and iron-based superconductors is an intriguing state of quantum matter[1, 2]. A major challenge in understanding superconductivity is the locally intertwined nature of various broken symmetry states, such as electronic nematic and smectic phases (that break rotational and translational symmetries of the electronic states), with superconductivity[3–7]. Manipulation and deliberate tuning of these broken symmetry states is a key objective toward understanding and controlling superconductivity.

Controlled strain, both uniaxial and biaxial, is a well-established technique to tune the collective electronic states in condensed matter systems[8–22]. This clean tuning, without the introduction of disorder through chemical doping, is commonly used in various kinds of experiments to tune bulk electronic properties[23–26]. For example, uniaxial pressure has proved to have an immense effect on superconductivity in $Sr_2RuO_4$[18] and cuprates[27] and on the structural, magnetic, and nematic phase transitions of iron-based superconductors[15, 19, 28, 29] and was recently demonstrated in tuning the topological states of $SmB_6$[24]. However, the use of strain in surface-sensitive techniques, such as STM and angle-resolved photoemission spectroscopy (ARPES), has been limited to in situ-grown thin films on mismatched substrates[26, 30]. The major challenge with applying strain to single crystals in surface-sensitive experiments is the need to cleave the strained samples in ultrahigh vacuum (UHV). In the last few years, an alternative direction has been to epoxy a thin sample on piezo stacks[9, 10, 15, 31] or on plates with different coefficients of thermal expansion[11, 32]. Yet in both cases, the magnitude of the applied strain is quite limited.

Here we demonstrate the use of a novel mechanical uniaxial-strain device that allows researchers to strain a sample (compressive strain) without constraints and simultaneously visualize its surface structure using STM (see **Figure 1**). As an example, we use single crystals of $Fe_{1+y}Te$, where $y = 0.10$, the parent compound of the iron chalcogenide superconductors (y is the excess iron concentration). Below $T_N$ = ~60 K, $Fe_{1+y}Te$ transitions from a high-temperature paramagnetic state into a low-temperature antiferromagnetic state with a bicollinear stripe magnetic order[26, 33, 34] (see **Figure 3A,B**). The magnetic transition is further accompanied by a structural transition from tetragonal to monoclinic[26, 35]. The in-plane AFM order forms detwinned domains with the spin structure pointing along the long b-direction of the orthorhombic structure[34]. By visualizing the AFM order with spin-polarized STM, we probe the bidirectional domain structure in unstrained $Fe_{1+y}Te$ samples and observe their transition into a single large domain under applied strain (see the schematic in **Figure 3C-E**). These experiments show the successful surface tuning of the single crystals using the uniaxial-strain device presented here, the cleaving of the sample, and the simultaneous imaging of its surface structure with the scanning tunneling microscope.

**Figure 1** shows the schematic drawings and pictures of the mechanical strain device. The U-shaped body is made of 416 grade stainless steel, which is stiff and has a low coefficient of thermal expansion $CTE \sim 9.9\ \mu m/m - °C$ as compared to $CTE \sim 17.3\ \mu m/m - °C$ for 304 grade stainless steel). The opening inside the U is 1mm and can be tuned smaller or large by a pair of micrometer screws located on the sides of the device. The uniaxial pressure is applied by the micrometer screw (1-72 corresponding to 72 rotations per inch). The sample, of size (1mm x 2mm x ~0.1mm) is mounted, with its long axis oriented along the b-axis of the sample, on top of the device using H20E conductive (silver) epoxy across the 1 mm gap. The two ends of the sample are then covered by H74F non-conductive epoxy for further reinforcement. Since uniaxial pressure is the tuning parameter in the experiment described here, it is imperative that the thermal stress generated from cooling down is not transferred directly to the sample. For this, we employ a series of Belleville spring disks. Using the working load of the Belleville spring disks of 67 N, and the deflection at working load of 50 μm, we calculate the spring constant for each disk as $k$ = 1.3 x $10^6$ N/m, which yields a total spring constant of $k$ = 1.625 x $10^5$ N/m for 4 pairs of springs in series. This ensures the thermal stress on the sample through cooling from room temperature to 4 K to be less than 0.05% for an applied strain of 1% and therefore negligible. In the experiment, we rotate the micrometer screw by 50° which corresponds to $Δx$ = 50 μm. The force applied on the sample through the springs can be calculated to be $F = kΔx$ = 8 N. The pressure is therefore $p = F/A$ = 8 N/(0.1 x $10^{-6}$ m²) = 0.08 GPa. For a Young's modulus of 70 Gpa for FeTe[36], the applied uniaxial pressure corresponds to 0.1% strain.

A major challenge in integrating the strain devices with the STM is the application of strain without breaking or introducing cracks in the sample. Test experiments on several samples of Bi-2212, $Sr_3Ru_2O_7$, and FeTe have shown that, depending on the sample thickness, the samples withstand strains of up to ~0.8-1.0 % corresponding to ~1GPa of applied pressure. No indications of cracks on the sample surface are observable below this value as seen visually by an optical microscope. Recent work following the same principles, has successfully demonstrated the application of ±1% strain on $Sr_2RuO_4$ [22]

Aluminum cleaving post is attached to strained sample perpendicular to the a-b cleaving plane using H74F non-conducting epoxy. The stain device with the sample and the post is then transferred, through the loading dock of the variable-temperature ultra-high vacuum STM, to the analysis chamber (see **Figure 2a**). All operations required to move the samples into and inside the STM are carried out using sets of arm manipulators. Using an arm manipulator, the aluminum post is knocked off in ultra-high vacuum at room temperature exposing a freshly cleaved surface. The device (with the strained sample) is then immediately transferred in situ with another set of manipulators to the STM chamber and into the microscope head (see **Figure 2b**), which had been cooled down to 9 K. All experiments are carried out at 9 K. The STM is maintained at low temperatures by liquid nitrogen and liquid helium, and the sample cools down for at least 12 h before being approached. This allows the sample and microscope temperature to reach thermal equilibrium. To isolate electric and acoustic noise, the STM is placed in an acoustic and radio frequency shielded room. The microscope head is further suspended from springs for optimized instrumental stability. The sample stage can be translated by several millimeters that enable access to different parts of the 1 mm strained samples.

STM topographs were taken in constant current mode with a setpoint bias of -12 meV applied to the sample and a setpoint current of -1.5 nA collected on the tip. Pt-Ir tips were used in all experiments. Tips were prepared prior to each experiment by field emission on a Cu (111) surface that had been treated with several rounds of sputtering and annealing. To achieve spin-polarized STM, the scanning tunneling microscope tip has to be coated with magnetic atoms, which can be quite challenging. In this case of studying $Fe_{1+y}Te$, the sample itself provides a simple means of achieving this. The excess irons ($y$ in $Fe_{1+y}Te$) are weakly bound on the cleaved surface. Scanning the tip at a low bias and with a high enough current exceeding a few nanoamperes brings the tip in close proximity to these Fe atoms and a few of those atoms can be picked up by the tip[37]. The other method that yields a spin-polarized tip is by the rapid decrease of the sample-tip separation until contact is made (on the location of excess iron concentration) as measured by a saturation current. During the process, the excess irons bond onto the tip. The successful preparation of a spin-polarized tip is revealed by the magnetic contrast in the topography, whose periodicity is twice that of the lattice constant of top tellurium atoms. This additional modulation is the antiferromagnetic order in the sample.

**Figure 4A** shows a 10 nm atomic-resolution topographical image on an unstrained $Fe_{1+y}Te$ single crystal with a nonmagnetic scanning tunneling microscope tip. The atomic structure seen corresponds to the Te atoms, which are exposed after cleaving the sample (see **Figure 3A**). The Fourier transform (FT) of the topography shows four sharp peaks at the corners of the image along the a- and b-directions, labeled $q_{Te}^a$ and $q_{Te}^b$, that correspond to the atomic Bragg peaks. The central broad peak in the FT corresponds to long-wavelength inhomogeneity, which is not relevant for the current study. **Figure 4C** shows another topograph of the same size as in **Figure 4A**, obtained with a magnetic tip. Unidirectional stripes with a periodicity of twice that of the lattice along the a-axis are observed. The FT of the topograph seen in **Figure 4D** shows, in addition to the Bragg peaks, a new pair of satellite peaks at $Q_{AFM1}$, corresponding to half the Bragg peak momenta and, therefore, twice the real space wavelength. The new structure corresponds to the AFM stripe order of the Fe atoms just below the surface.

On this unstrained sample, it is not difficult to observe twin domain boundaries where the crystal structure with the long b-axis and the accompanying AFM stripe order rotate 90°. **Figure 4E** shows a 25 nm spin-polarized topograph of an AFM twin domain boundary. The FT of the image now shows two pairs of AFM order (highlighted by green and yellow circles). Each magnetic domain contributes to only one pair of the $Q_{AFM}$ peaks in the FT. To visualize this clearly, we Fourier-filtered each pair of AFM peaks and inversed FT back to real space. The results are shown in **Figure 4G,H** highlighting the two unidirectional stripe domains.

Thus, we studied the domain structure and boundaries on the surface on a large scale. **Figure 5A**, **Figure 6A**, and **Figure 7A** display large-scale topographs on three different unstrained samples spanning a total region of slightly over 0.75 μm x 0.75 μm. Several smaller zoomed-in topographs are also shown to highlight the stripe structure. The topographs are taken with a high spatial resolution (1024 x 1024 pixels per 0.25 μm$^2$) to allow the Fourier filtering and inverse Fourier transform analysis on the large scale. The corresponding domain structures and boundaries are displayed in **Figure 5B**, **Figure 6C**, and **Figure 7H**. Overall, several alternating stripe domains are observed covering the overall equal areas, as expected for these unstrained samples. It is important to note that on this large scale the surface is overall atomically flat, yet a few different structural irregularities, such as line defects (**Figure 5A**) and atomic steps (**Figure 7A**), can be observed. The stripe domains are not affected by these irregularities.

From here, we moved on to the strained sample. **Figure 8** shows a large-scale topograph, spanning a total region of ~1.75 μm x 0.75 μm, which is more than twice the total area spanned in the unstrained samples shown in **Figure 5**, **Figure 6**, and **Figure 7**. In stark contrast, the FT for each topograph shows only one pair of AFM peaks indicating only a single domain on this strained sample. This can further be visualized by the Fourier-filtered iFT analysis confirming the single stripe domain over the entire area. Once again, the unidirectional stripe order is not to be affected by the different surface irregularities in this strained sample.

The success of this technique lies in the careful execution of the correct alignment of the sample across the 1 mm gap and application of the strain on the sample without breaking or bending it. Another important consideration is the cleaving process, which allows the exposure of a clean flat surface. This is a random process and works best for materials that cleave easily. A last consideration is having a sharp tip that yields atomic resolution and can pick up some excess iron atoms to achieve magnetic contrast.

In conclusion, the experiments and analysis described here successfully demonstrate the incorporation of our strain device with STM, providing a new tuning parameter that can be invaluable in the study of competing orders in correlated electron systems. The advantage of the current device is the wide range of positive and negative strain that can be applied to the sample. This demonstration may impact other spectroscopic experiments such as ARPES.

**REFERENCES:**

1. Paglione, J. & Greene, R. L. High-temperature superconductivity in iron-based materials. *Nature Physics.* **6** (9), 645, (2010).
2. Keimer, B., Kivelson, S. A., Norman, M. R., Uchida, S. & Zaanen, J. From quantum matter to high-temperature superconductivity in copper oxides. *Nature.* **518** 179-186, (2015).
3. Anderson, P. W. Physics: The opening to complexity. *Proceedings of the National Academy of Sciences.* **92** (15), 6653-6654, (1995).
4. Dagotto, E. Complexity in strongly correlated electronic systems. *Science.***309** , 257-262, (2005).
5. Davis, J. S. & Lee, D.-H. Concepts relating magnetic interactions, intertwined electronic orders, and strongly correlated superconductivity. *Proceedings of the National Academy of Sciences.* **110** (44), 17623-17630, (2013).
6. Fernandes, R., Chubukov, A. & Schmalian, J. What drives nematic order in iron-based superconductors? *Nature Physics.* **10** (2), 97, (2014).
7. Fradkin, E., Kivelson, S. A. & Tranquada, J. M. Colloquium: Theory of intertwined orders in high temperature superconductors. *Reviews of Modern Physics.* **87** (2), 457, (2015).
8. Stillwell, E., Skove, M. & Davis, J. Two``Whisker''Straining Devices Suitable for Low Temperatures. *Review of Scientific Instruments.* **39** (2), 155-157, (1968).
9. Shayegan, M. *et al.* Low-temperature, in situ tunable, uniaxial stress measurements in semiconductors using a piezoelectric actuator. *Applied physics letters.* **83** (25), 5235-5237, (2003).
10. Yim, C. M. *et al.* Discovery of a strain-stabilised smectic electronic order in LiFeAs. *Nature communications.* **9** (1), 2602, (2018).
11. Gao, S. *et al.* Atomic-scale strain manipulation of a charge density wave. *Proceedings of the National Academy of Sciences.* 115 (27) 6986-6990, (2018).
12. Jiang, J. *et al.* Distinct in-plane resistivity anisotropy in a detwinned FeTe single crystal: Evidence for a Hund's metal. *Physical Review B.* **88** (11), 115130, (2013).
13. Zhang, Y. *et al.* Symmetry breaking via orbital-dependent reconstruction of electronic structure in detwinned NaFeAs. *Physical Review B.* **85** (8), 085121, (2012).
14. Watson, M. D., Haghighirad, A. A., Rhodes, L. C., Hoesch, M. & Kim, T. K. Electronic anisotropies revealed by detwinned angle-resolved photo-emission spectroscopy measurements of FeSe. *New Journal of Physics.* **19** (10), 103021, (2017).
15. Chu, J.-H., Kuo, H.-H., Analytis, J. G. & Fisher, I. R. Divergent nematic susceptibility in an iron arsenide superconductor. *Science.* **337** (6095), 710-712, (2012).
16. Song, Y. *et al.* Uniaxial pressure effect on structural and magnetic phase transitions in NaFeAs and its comparison with as-grown and annealed BaFe2As2. *Physical Review B.* **87** (18), 184511, (2013).
17. M. P. Allan *et al.* Anisotropic impurity states, quasiparticle scattering and nematic transport in underdoped Ca(Fe1–xCox)2As2. *Nature Physics.* **9** (4), 220-224, (2013).
18. Hicks, C. W. *et al.* Strong increase of Tc of Sr2RuO4 under both tensile and compressive strain. *Science.* **344** (6181), 283-285, (2014).

19. Hicks, C. W., Barber, M. E., Edkins, S. D., Brodsky, D. O. & Mackenzie, A. P. Piezoelectric-based apparatus for strain tuning. *Review of Scientific Instruments.* **85** (6), 065003, (2014)
20. Gannon, L. *et al.* A device for the application of uniaxial strain to single crystal samples for use in synchrotron radiation experiments. *Review of Scientific Instruments.* **86** (10), 103904, (2015).
21. Kretzschmar, F. *et al.* Critical spin fluctuations and the origin of nematic order in Ba(Fe1–xCox)2As 2. *Nature Physics.* **12** (6), 560, (2016).
22. Steppke, A. *et al.* Strong peak in T c of Sr2RuO4 under uniaxial pressure. *Science.* **355** (6321), 133, (2017).
23. Iida, K. *et al.* Strong T c dependence for strained epitaxial Ba(Fe1-xCox)2As2 thin films. *Applied physics letters.* **95** (19), 192501, (2009).
24. Stern, A., Dzero, M., Galitski, V., Fisk, Z. & Xia, J. Surface-dominated conduction up to 240 K in the Kondo insulator SmB6 under strain. *Nature materials.* **16** (7), 708-711, (2017).
25. Iida, K. *et al.* Hall-plot of the phase diagram for Ba(Fe1–xCox)2As2. *Scientific reports.* **6** 28390, (2016).
26. Hänke, T. *et al.* Reorientation of the diagonal double-stripe spin structure at Fe1+yTe bulk and thin-film surfaces. *Nature communications.* **8** 13939, (2017).
27. Takeshita, N., Sasagawa, T., Sugioka, T., Tokura, Y. & Takagi, H. J. Gigantic anisotropic uniaxial pressure effect on superconductivity within the CuO2 plane of La1.64Eu0.2Sr0.16CuO4: Strain control of stripe criticality. *Journal of the Physical Society of Japan.* **73** (5), 1123-1126, (2004).
28. Kuo, H.-H., Shapiro, M. C., Riggs, S. C. & Fisher, I. R. Measurement of the elastoresistivity coefficients of the underdoped iron arsenide Ba(Fe0.975Co0.025)2As2. *Physical Review B.* **88** (8), 085113, (2013).
29. He, M. *et al.* Dichotomy between in-plane magnetic susceptibility and resistivity anisotropies in extremely strained BaFe2As2. *Nature communications.* **8** (1), 504, (2017).
30. Engelmann, J. *et al.* Strain induced superconductivity in the parent compound BaFe2As2. *Nature communications.* **4** (2877), 2877, (2013).
31. Denise, A., Berger, N. Temperature Driven Topological Switch in 1T'-MoTe2 and Strain Induced Nematicity in NaFeAs (2018).
32. Böhmer, A. *et al.* Effect of biaxial strain on the phase transitions of Ca(Fe1–xCox)2As2. *Physical review letters.* **118** (10), 107002, (2017).
33. Bao, W. *et al.* Tunable (δ π, δ π)-type antiferromagnetic order in α-Fe(Te,Se) superconductors. *Physical review letters.* **102** (24), 247001, (2009).
34. Koz, C., Rößler, S., Tsirlin, A. A., Wirth, S. & Schwarz, U. Low-temperature phase diagram of Fe1+yTe studied using x-ray diffraction. *J Physical Review B.* **88** (9), 094509, (2013).
35. Enayat, M. *et al.* Real-space imaging of the atomic-scale magnetic structure of Fe1+yTe. *Science.* **345** (6197), 653-656, (2014).
36. Chandra, S., & Islam, A. K. M. A. Elastic and electronic properties of PbO-type FeSe1-xTex (x= 0-1.0): A first-principles study. *ArXiv preprint arXiv:1008.1448*, (2010).
37. Singh, U. R., Aluru, R., Liu, Y., Lin, C. & Wahl, P. Preparation of magnetic tips for spin-polarized scanning tunneling microscopy on Fe1+yTe. *Physical Review B.* **91** (16), 161111, (2015).

**ACKNOWLEDGMENTS:**
P.A. acknowledges support from the U.S. National Science Foundation (NSF) CAREER under award No. DMR-1654482. Material synthesis was carried out with the support of the Polish National Science Centre grant No 2011/01/B/ST3/00425.

**FIGURE LEGENDS:**

**Figure 1: Strain device.** (**A**) Schematic of the strain device. The U-shaped device has two micrometer screws for the (1) compression and (2) expansion of the device's gap area. The sample can be confined inside the gap as shown in figure panels **A** and **C** or on top of the gap as shown in figure panels **A** and **B**. A combination of H20E and H74F epoxies are applied to the sample and cured at 100 °C. Once the epoxy on the sample is cured, a post of about the same surface area as that of the sample is epoxied onto the sample's surface using H74F. (**B**) The actual setup of the strain device, with a top view, front view, and a zoom-in of the sample. The device is screwed to a sample holder that slides into the microscope head. A contact is created by using conductive epoxy from the device to the sample plate. The transfer of pressure is enabled using a screw and a series of Belleville spring disks. The last panel of **B** shows the strain device set up, ready to be moved into the UHV analysis chamber. (**C**) An alternative method is to have a sample inside the gap of the strain device. In the two middle panels of **C**, a second unstrained sample is epoxied on the device for reference.

**Figure 2: Scanning tunneling microscope setup.** (**A**) The scanning tunneling microscope setup. The microscope is placed in an acoustic chamber, which is shielded from radio-frequency (RF) noise. (**B**) The microscope head with a bare sample holder. The Pt/Ir tip is visible. The sample stage can be moved by a set of piezo actuators so that the sample is right above the tip. (**C**) The microscope head is placed inside two radiation shields.

**Figure 3: $Fe_{1+y}Te$ crystal structure.** (**A**) The crystal structure of FeTe with the top layer showing the tellurium atoms. The red dotted lines outline the three unit cells. (**B**) A real-space schematic illustration of the atomic unit cell (red solid line) and magnetic structure (black solid line) of FeTe. The magnetic wavevector $\lambda_{afm}$ is twice the atomic distance between Te-Te atoms. The arrows on the Fe atoms indicate the spin orientations. (**C**) Schematic diagram illustrating the AFM twin domains that form when cooling, through the structural transition from tetragonal to monoclinic at ~60 to 70 K, with an equal population of the two domains. (**D**) The response of the detwinning process, when an appreciable amount of strain is applied along the b-axis (black arrows) with one domain enhanced (red) and the other domain diminished (blue). (**E**) A complete detwinned domain, which leaves only one single domain. (**F–H**) The FT of the real space in panels **C–E**. The $Q_{AFM1}$ peaks correspond to the red real-space domains, and the $Q_{AFM2}$ peaks correspond to the blue domains. The lattice Bragg peaks are denoted as black dots at the corners of the image.

**Figure 4: Unidirectional modulation from unstrained $Fe_{1+y}Te$.** (**A**) A 10 nm x 10 nm topograph of the atomic lattice structure of $Fe_{1+y}Te$ with no magnetic contrast. (**B**) The FT of panel **A**, showing the Bragg peaks at the corners of the images (black circles). (**C**) A 10 nm x 10 nm topograph of the magnetic structure of $Fe_{1+y}Te$, measured using a spin-polarized tip. The unidirectional stripes across the a-axis correspond to peaks appearing at $Q_{AFM1} = q_{Te}^a/2$ in the FT, as shown in panel **D**. (**E**) A 25 nm x 25 nm topographical image across a twin domain boundary. (**F**) FT of panel **E**, showing the two sets of peaks $Q_{AFM1}$ and $Q_{AFM2}$. (**G**) Inverse Fourier transform (iFT) of the $Q_{AFM1}$ peaks from panel **F**. The red color corresponds to the high intensity of the $Q_{AFM1}$ peaks. (**H**) iFT of the $Q_{AFM2}$ peaks from panel **F**. The domain boundary is clearly distinct from the images shown in panels **G** and **H**. The inverse Fourier filtering method has been used in subsequent figures to identify the different domains.

**Figure 5: Imaging twin domains in unstrained Fe$_{1+y}$Te.** (**A**) A 0.75 μm x 0.25 μm topographical image showing twin boundaries. The data was acquired in three adjacent topographical images, each 0.25 μm x 0.25 μm. (**B**) Using iFT, the domain boundaries are distinctly evident. (**C–E**) Zoom-ins of the images marked with an (X) and a yellow-colored dotted box are shown with highlighted, dotted, colored boxes around the boundaries.

**Figure 6: Imaging multiple domains from unstrained Fe$_{1+y}$Te.** (**A**) A 0.10 μm x 0.10 μm topographical image of an unstrained Fe$_{1+y}$Te. (**B**) The FT of panel **A**, which shows peaks in both directions, namely $Q_{AFM1}$ and $Q_{AFM2}$. (**C**) The iFT image of panel **A**, indicating the different domains. (**D** and **E**) Zoom-ins of the highlighted yellow- and orange-dotted boxes in panel **A**.

**Figure 7: Imaging twin domains from unstrained Fe$_{1+y}$Te.** (**A**) Topographical images spanning an area of 0.75 μm x 0.5 μm. (**B–D**) Line cuts of the topograph taken across the black, purple, and green arrows in panel **A**. (**E–G**) Zoom-in of the areas highlighted in the green, brown, and yellow (X) marks in panel **A**. (**H**) iFT of panel **A**, showing the twin domains. The white-dotted lines are the step edges/boundaries. The domains are unaffected by these structural features.

**Figure 8: Imaging detwinned domains in strained Fe$_{1+y}$Te.** (**A**) A large 1.750 μm x 0.50 μm topography taken on a strained Fe$_{1+y}$Te sample. (**B** and **C**) The FT of the two largest (0.50 μm x 0.50 μm) single topographs acquiring on one pair of AFM peaks in one direction. (**D**) The Fourier-filtering and iFT process is applied to the images in panel **A**, which shows only a single domain as expected. The dotted line in panel **D** is a step which does not affect the unidirectional domain. (**E**) A zoom-in of the highlighted region in the yellow (X) showing unidirectional stripes. (**F**) A zoom-in of panel **E**, also showing clearly the unidirectional stripes of the detwinned sample. (**G**) The FT of panel **E**. The AFM peaks appear only in one direction, which agrees with the real-space structure in panel **E**.

**Figure 1**

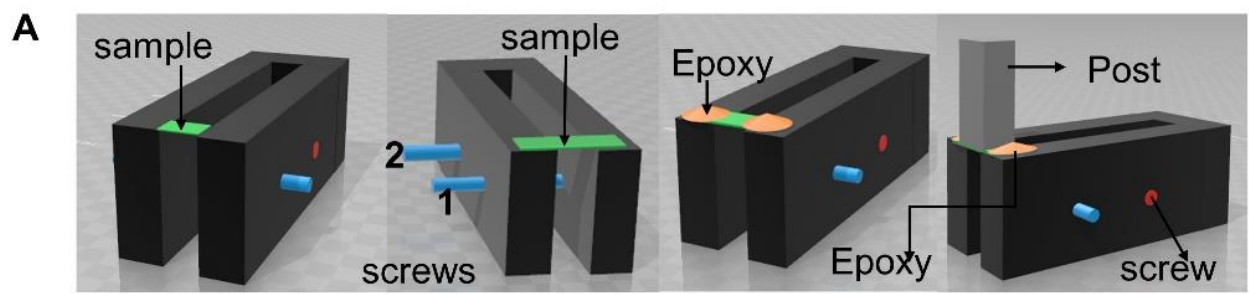

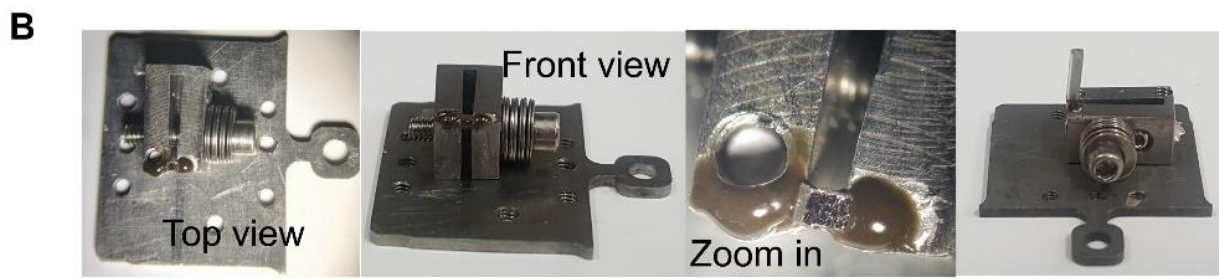

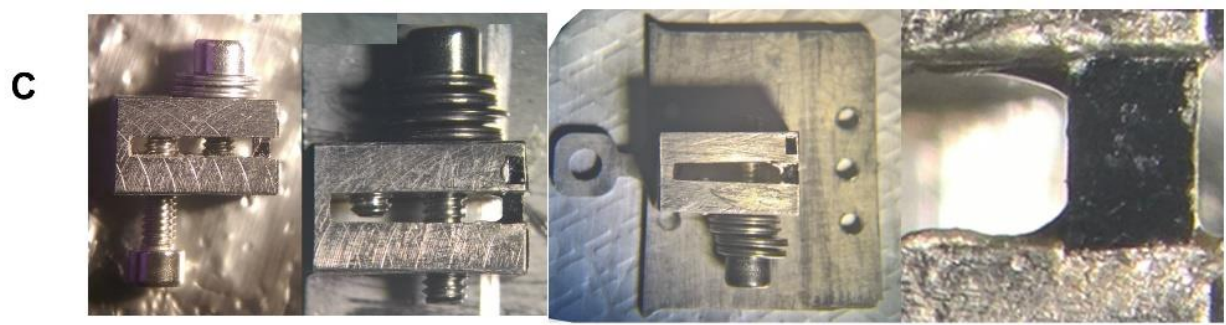

**Figure 2**

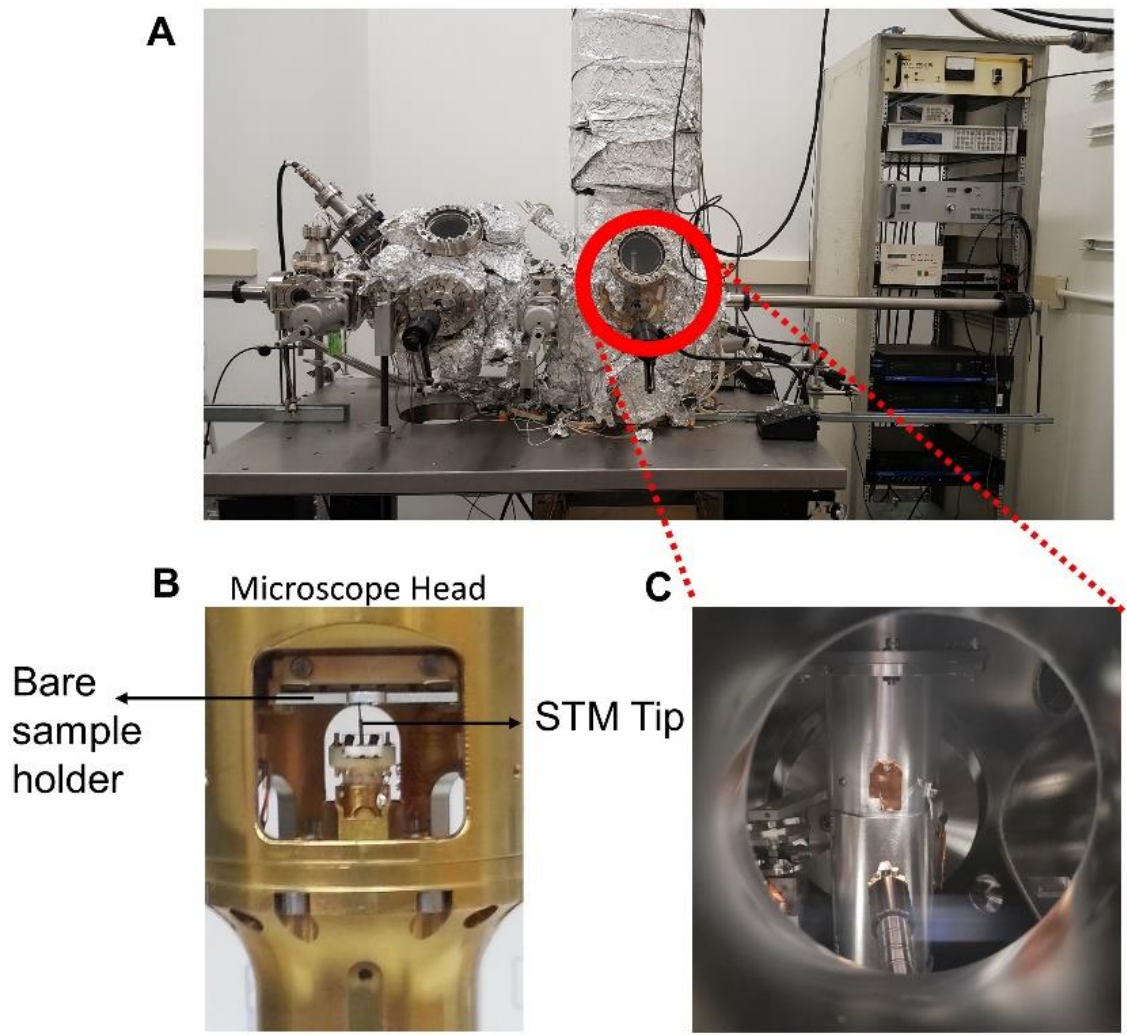

Figure 3

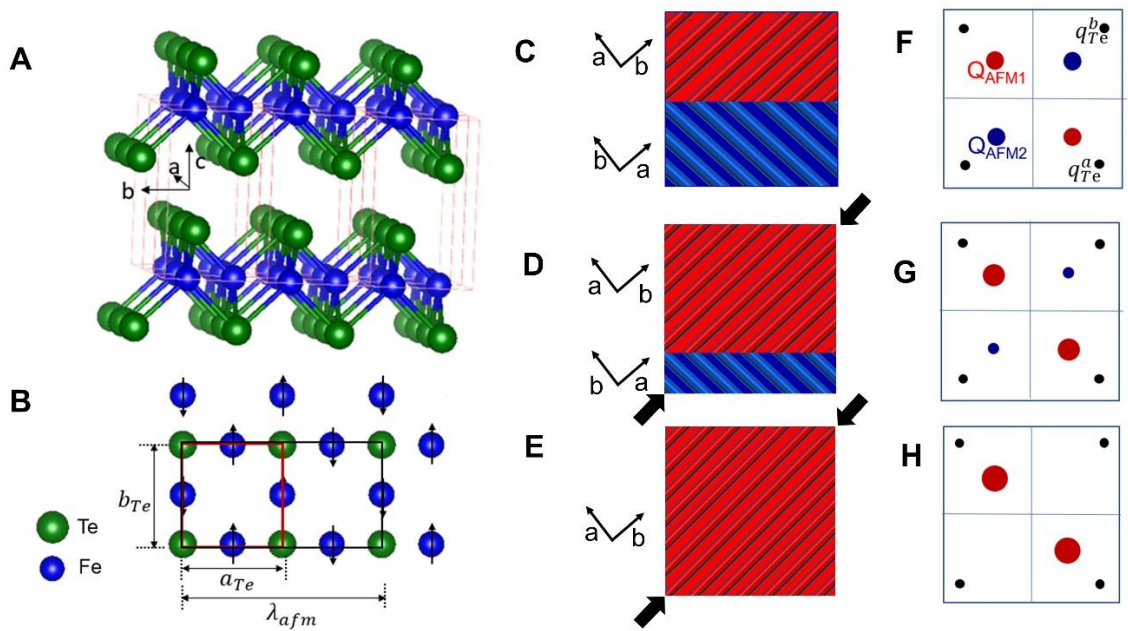

**Figure 4**

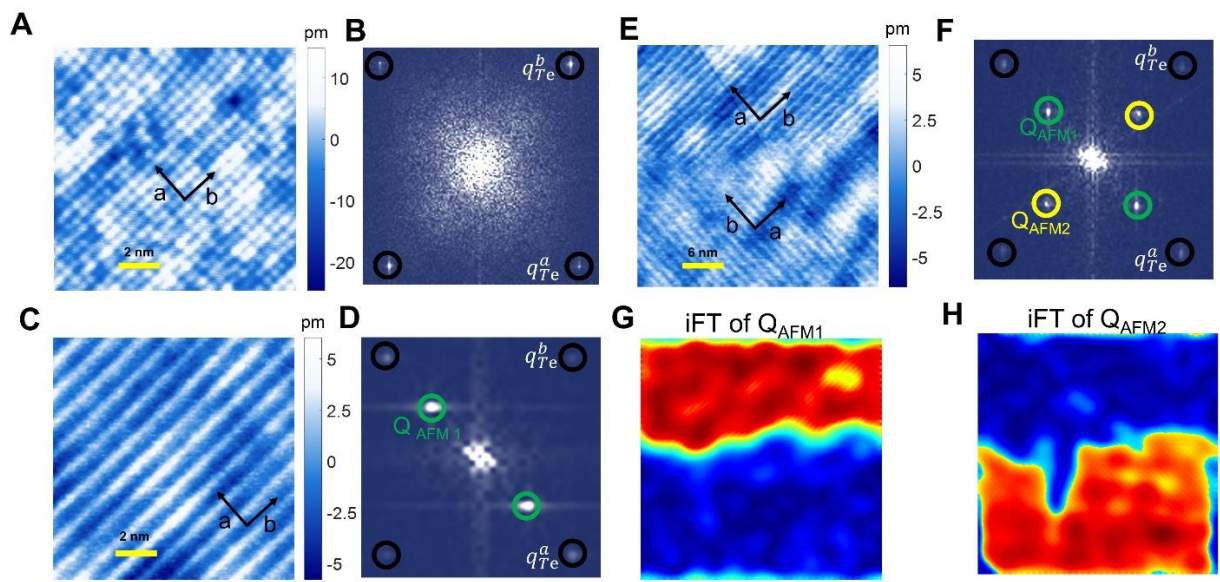

**Figure 5**

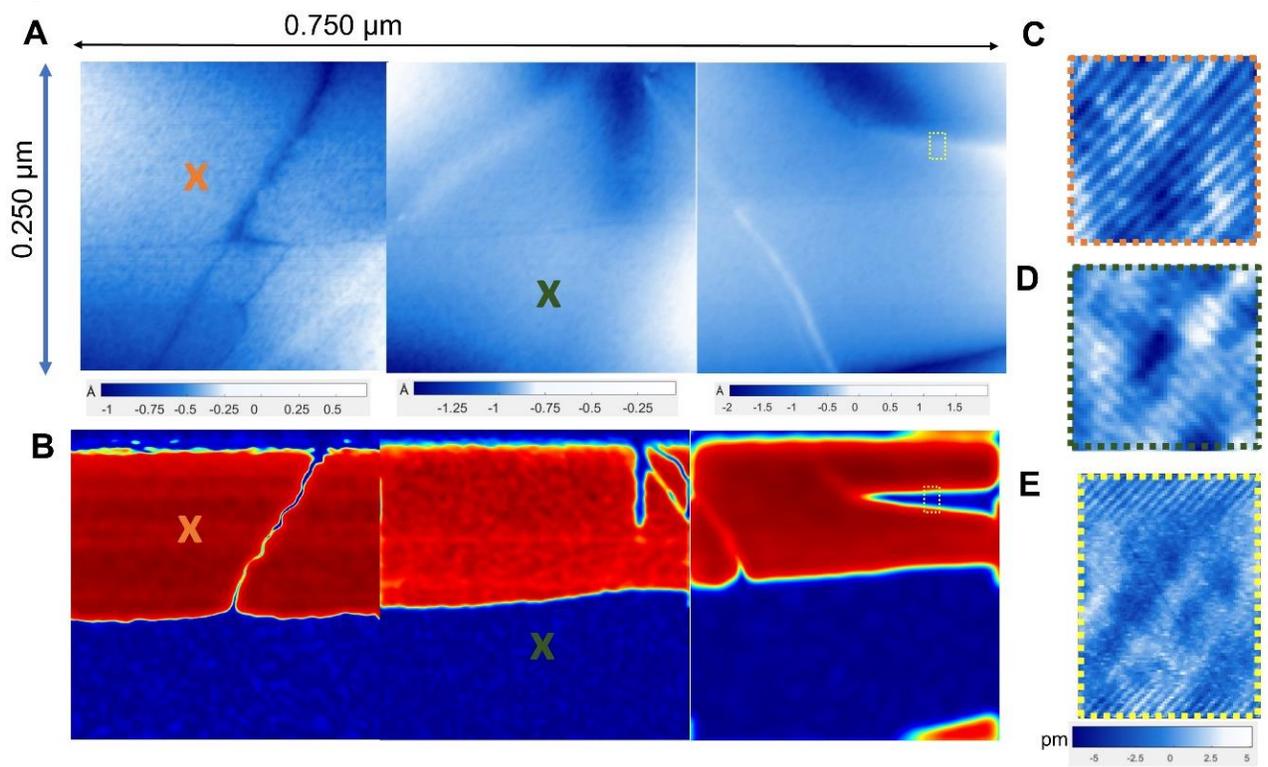

**Figure 6**

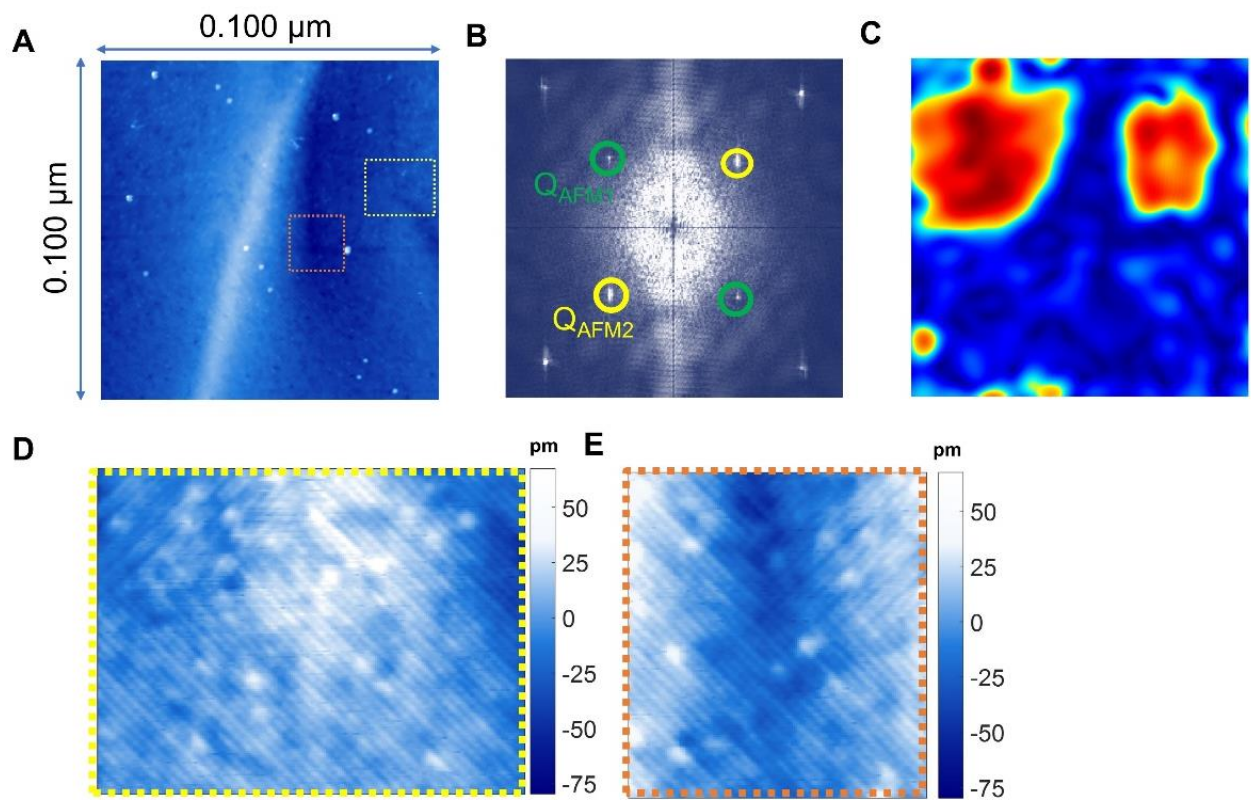

**Figure 7**

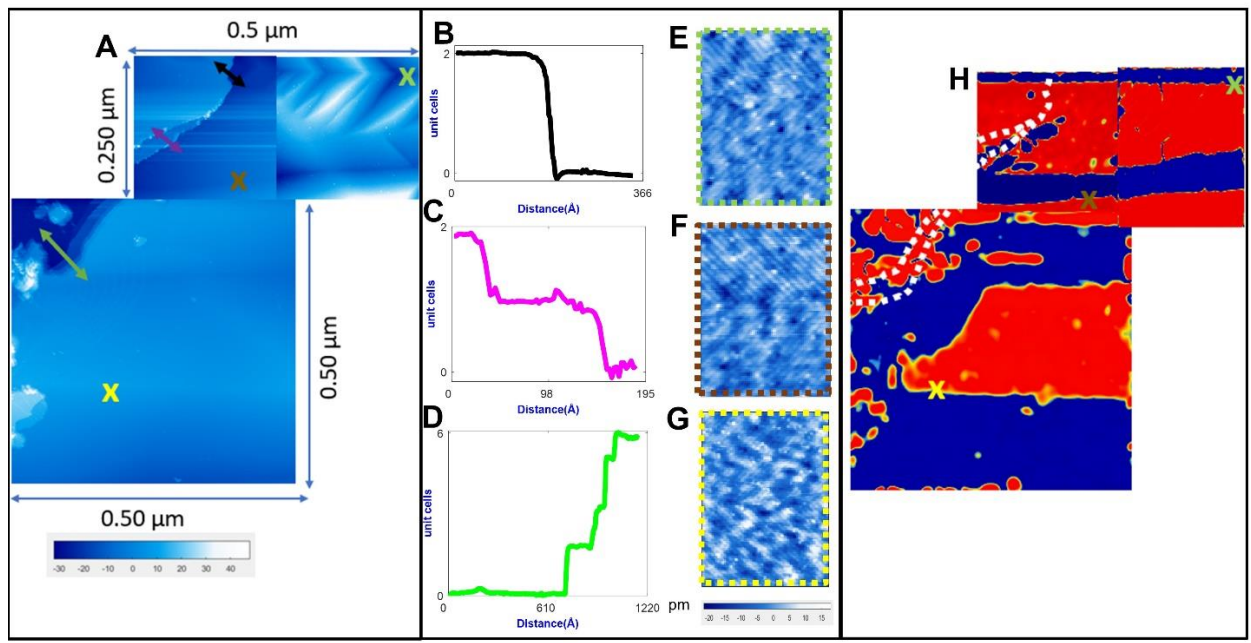

**Figure 8**

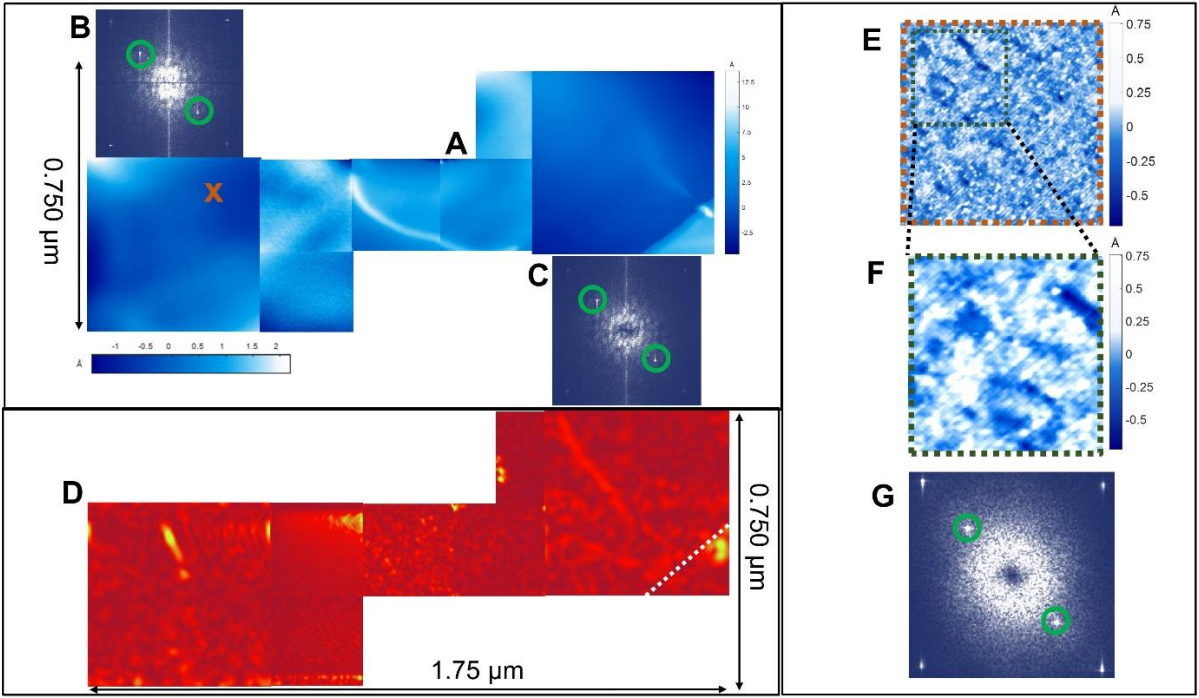